\documentclass[conference,hidelinks]{article}

\usepackage{arxiv}

\usepackage{cite}

\usepackage[caption=false,font=normalsize,labelfont=sf,textfont=sf]{subfig}

\usepackage{xfrac}
\usepackage{todonotes}
\usepackage{booktabs}
\usepackage{listings}
\usepackage{xcolor}
\lstdefinestyle{codeblockstyle}{language=C++,
    basicstyle=\scriptsize\bfseries\ttfamily,
    keywordstyle=\color{blue}\bfseries,
    stringstyle=\color{red}\bfseries,
    commentstyle=\color{purple}\bfseries
}
\lstdefinestyle{inlinestyle}{language=C++,
    basicstyle=\small\bfseries\ttfamily,
    keywordstyle=\color{blue}\bfseries,
    stringstyle=\color{red}\bfseries,
    commentstyle=\color{purple}\bfseries
}
\lstdefinestyle{inlinestyle-no-language}{
    basicstyle=\small\bfseries\ttfamily,
    keywordstyle=\color{blue}\bfseries,
    stringstyle=\color{red}\bfseries,
    commentstyle=\color{purple}\bfseries
}

\definecolor{plot1}{HTML}{003f5c}
\definecolor{plot2}{HTML}{bc5090}
\definecolor{plot3}{HTML}{ffa600}

\usepackage{hyperref}
\usepackage{booktabs}

\usepackage{tikz}
\usepackage{pgfplots}

\newcommand{\orcid}[1]{\href{https://orcid.org/#1}{\includegraphics[height=10pt]{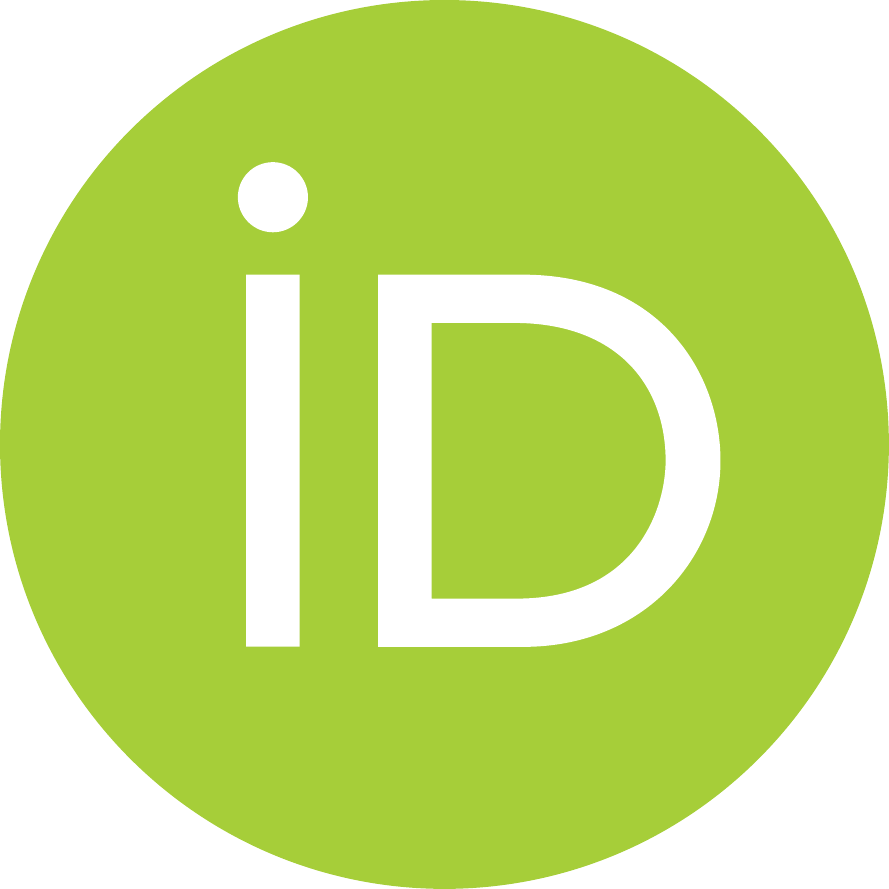}}}

\usepackage{aas_macros}

\begin{document}
%\bstctlcite{IEEEexample:BSTcontrol}
%
% paper title
% Titles are generally capitalized except for words such as a, an, and, as,
% at, but, by, for, in, nor, of, on, or, the, to and up, which are usually
% not capitalized unless they are the first or last word of the title.
% Linebreaks \\ can be used within to get better formatting as desired.
% Do not put math or special symbols in the title.
\title{Octo-Tiger's New Hydro Module and Performance Using HPX+CUDA on ORNL's Summit}

% author names and affiliations
% use a multiple column layout for up to three different
% affiliations
\author{
Patrick Diehl\orcid{0000-0003-3922-8419} \\
LSU Center for Computation \& Technology, Louisiana State University,
Baton Rouge, LA, 70803 U.S.A\\
Email: patrickdiehl@lsu.edu
\AND Gregor Dai\ss\\
IPVS, University of Stuttgart,  Stuttgart, 70174 Stuttgart, Germany
\AND Dominic Marcello\\
Department of Physics and Astronomy, Louisiana State University,
Baton Rouge, LA, 70803 U.S.A.
\AND Kevin 
Huck\orcid{0000-0001-7064-8417}\\
\textit{OACISS}, University of Oregon, Eugene, OR, U.S.A.
\AND Sagiv Shiber\orcid{0000-0001-6107-0887}\\
Department of Physics and Astronomy, Louisiana State University,
Baton Rouge, LA, 70803 U.S.A.
\AND Hartmut 
Kaiser\orcid{0000-0002-8712-2806} \\
LSU Center for Computation \& Technology, Louisiana State University,
Baton Rouge, LA, 70803 U.S.A
\AND Juhan Frank, Geoffrey C. Clayton\orcid{0000-0002-0141-7436}\\
Department of Physics and Astronomy, Louisiana State University,
Baton Rouge, LA, 70803 U.S.A.
\AND  Dirk Pfl\"uger\orcid{0000-0002-4360-0212} \\
IPVS, University of Stuttgart,  Stuttgart, 70174 Stuttgart, Germany}

\maketitle

% As a general rule, do not put math, special symbols or citations
% in the abstract
\begin{abstract}
Octo-Tiger is a code for modeling three-dimensional self-gravitating astrophysical fluids. It was particularly designed for the study of dynamical mass transfer between interacting binary stars. Octo-Tiger is parallelized for distributed systems using the asynchronous many-task runtime system, the C++ standard library for parallelism and concurrency (HPX) and utilizes CUDA for its gravity solver. Recently, we have remodeled Octo-Tiger's hydro solver to use a three-dimensional reconstruction scheme. In addition, we have ported the hydro solver to GPU using CUDA kernels. We present scaling results for the new hydro kernels on ORNL's Summit machine using a Sedov-Taylor blast wave problem. We also compare Octo-Tiger's new hydro scheme with its old hydro scheme, using a rotating star as a test problem.
\end{abstract}

% no keywords

% For peer review papers, you can put extra information on the cover
% page as needed:
% \ifCLASSOPTIONpeerreview
% \begin{center} \bfseries EDICS Category: 3-BBND \end{center}
% \fi
%
% For peerreview papers, this IEEEtran command inserts a page break and
% creates the second title. It will be ignored for other modes.
%\IEEEpeerreviewmaketitle

\clearpage
\newpage
\twocolumn

\section{Introduction}
% no \IEEEPARstart
Octo-Tiger is an astrophysics finite volume hydrodynamic code for simulating the evolution of stellar systems~\cite{marcello2021octo}. Octo-Tiger~ consists of several modules, \emph{e.g.}\ hydro, gravity, and radiation. The gravity is solved based on the fast multipole method using adaptive octrees. The hydro module solves the mass, momentum and energy equations of an inviscid fluid in a rotating frame of reference, which reduces numerical viscosity effects.
Recently, we improved the accuracy of the hydro module by including a full three-dimensional reconstruction technique (see a thorough introduction of this technique in~\cite{marcello2021octo}). With the fully three-dimensional reconstruction, the hydro module became the hotspot of the application. Here, we present and test its initial GPU implementation. Our radiation module, still in the testing phase, uses an explicit transport scheme with the reduced speed of light approximation, coupled to an implicit scheme for the radiation-hydro coupling terms, in a manner similar to Skinner et al.~\cite{Skinner2013}.

To validate the theoretical claim that the full three-dimensional reconstruction technique results in more accuracy, a rotating star simulation using the old and new hydro modules with the same gravity module were executed. The error and convergence of both methods is compared to validate the theoretical claim with numerical results, see Section~\ref{sec:astro:results}. However, this paper focuses on the task-based execution using adaptive mesh refinement, resulting in some irregular parallelism. The task-based approach helps us with properly parallelizing the tree-traversals. As we strive for the lowest time per timestep possible, this in turn means we have to process millions of cells %thousands of subgrids 
in sub-second runtimes. This means we have a task-graph of extremely short running compute kernels mixed with the communication and data transfers.

We are revisiting the performance of the gravity module and studying the performance of the new hydro module on ORNL's Summit. Octo-Tiger's scaling capabilities have been previously shown: NERSC's Cori~\cite{heller2019harnessing} and on CSCS Piz Daint~\cite{daiss2019piz}, however, in these measurements an older version of the hydro module was used. We have experience running Octo-Tiger and the C++ standard library for parallelism and concurrency (HPX)~\cite{Kaiser2020} on x86 systems and CRAY based systems, but not much previous experience with distributed runs on IBM\textregistered~Power9\texttrademark\ systems.

First, the hydro module for the Sedov-Taylor blast wave is studied. Second, a rotating star for the combination of the hydro and gravity module is simulated. For both problems, we show the node level scaling for CPU and CPU+GPU runs on a single node. Note that due to the different implementations of the hydro kernels, especially the more computationally intense reconstruction of the fluxes in the new implementation, we can not directly compare the scaling results.

In addition, analyzing such large task graphs can be rather challenging, see Figure~\ref{fig:taskgraphs}. This is the first time we employ APEX with CUDA support to get combined profiling of the CPU and GPU tasks. CPU-only profiling with APEX has been shown in~\cite{diehl2021performance}.

The paper is structured as follows: Section~\ref{sec:related:work} covers the related work. Section~\ref{sec:software:framework} sketches the software framework. Section~\ref{sec:octo} introduces Octo-Tiger's new hydro module and its GPU acceleration. Section~\ref{sec:performance} shows the node level and distributed scaling of Octo-Tiger on Summit. Section~\ref{sec:astro:results} compares the accuracy of the new three-dimensional full reconstruct of the hydro kernel with the previous kernel. Finally, Section~\ref{sec:sonclusion} concludes the paper.

%%%%%%%%%%%%%%%%%%%%%%%%%%%%%%%%%%%
\section{Related work}
\label{sec:related:work}
There are many astrophysics codes which combine hydrodynamic and gravity solvers for the simulation of astrophysical fluids. Here, however, we are focusing on those which have two additional properties that Octo-Tiger has: \textit{1)} They are accelerated by an asynchronous many-task system (AMT) and \textit{2)} They use adaptive grid refinement. ChaNGa (Charm N-body Gravity solver)~\cite{4536319} performs collisionless N-body simulations for cosmological simulations or simulations of isolated stellar systems. A moving-mesh hydrodynamic solver was added to ChaNGa~\cite{Chang2017} together with the implementation of multiple time-steps techniques to form the code MANGA~\cite{Prust2019}, suitable for simulating interacting binary stars. Enzo-E \slash\ Cello (formerly Enzo-P)~\cite{10.5555/2462077.2462081}, which is currently under active development, is designed for astrophysics simulations, including star formation and cosmology applications. Cello provides the AMR feature within Enzo-E. Both of these codes use the AMT Charm++~\cite{10.1145/165854.165874}. Another AMR-based code is Castro~\cite{Almgren2020}, part of the AMReX Astrophysics suite utilizing the more traditional MPI+X approach. The Athena++ code, a C++ rewrite of the magneto-hydrodynamic code Athena C, implements an adaptive mesh refinement and uses MPI+OpenMP for its parallelization \cite{Stone2020}. A GPU-accelerated version of Athena++, K-Athena, was refactored using Kokkos to achieve better performance and portability \cite{Grete2019}. All these codes attempt to exploit high abstraction programming for the parallelization of their code to display scaling on exascale supercomputers.
For example, Charm++ and the AMT used by Octo-Tiger, HPX, have very similar programming models. From an application developer perspective, HPX can be seen as an abstraction to C++ and Charm++ more as a standalone library~\cite{tbaa20-report}. According to this survey~\cite{thoman2018taxonomy} HPX has the highest technical readiness. Two of the codes, K-Athena and Castro, have recently reported their scaling and performance on OLCF's Summit~\cite{Grete2019,Katz2020}. We aim to report Octo-Tiger's performance on Summit as well, in particular after upgrading the hydro solver and porting it to GPU CUDA kernels. Since two of the main functionalities of the code, the gravity and hydro solvers, can be executed on GPUs, it is interesting to study the scaling on numerous GPUs. Although a direct comparison between the performance of codes is not trivial, a simple basic measurement of interest is the number of cells (zones) updated per second (or per microseconds). Castro reported a value of $130$ zones/$\mu$seconds on one Summit node~\cite{Katz2020}, while K-Athena reported a peak value of $>100$ zones/$\mu$seconds~\cite{Grete2019}.

%%%%%%%%%%%%%%%%%%%%%%%%%%%%%%%%%%%
\section{Software framework}
\label{sec:software:framework}

%%%%%%%%%%%%%%%%%%%%%%%%%%%%%%%%%%%
\subsection{C++ standard library for parallelism and concurrency}
\label{sec:hpx}
HPX is the C++ standard library for parallelism and concurrency. It exposes an API that fully conforms to the recent C++ standards~\cite{cxx11_standard,cxx14_standard,cxx17_standard,cxx20_standard} on top of an asynchronous many-task runtime system (AMT). It has been described in detail in other publications, such as~\cite{hpx_pgas_2014,Kaiser:2015:HPL:2832241.2832244,hartmut_kaiser_2021_598202,Heller2016,Kaiser2020}. In the context of this paper, HPX has been used for two purposes. a) to coordinate the asynchronous execution of a multitude of heterogeneous tasks (both on CPUs and GPUs), thus managing local and distributed parallelism while observing all necessary data dependencies, and b) as the parallelization infrastructure for executing CUDA-kernels on the CPUs via the asynchronous HPX backend.

%%%%%%%%%%%%%%%%%%%%%%%%%%%%%%%%%%%
\subsection{APEX}

%APEX overview
%new CUDA/CUPTI/NVML support, callbacks, activity and monitoring
%runtime, driver API options, detailed counter options, memory allocation tracking
APEX~\cite{huck2015autonomic} is a
performance measurement library for distributed, asynchronous multitasking
systems. It provides lightweight measurements without perturbing high
concurrency through synchronous and asynchronous interfaces.
To support performance measurement in systems that employ
%operating system- or
user-level threading, APEX uses a dependency chain in addition to the call stack to
produce traces and task dependency graphs.
The synchronous APEX instrumentation application programming interface (API) can be used to add instrumentation to a given
run time and includes support for timers and counters.
%To support C++ threads on Linux systems, the underlying POSIX threads are automatically instrumented by using a preloaded shared object library that intercepts and wraps pthread calls in the application.
The NVIDIA CUDA Profiling Tools Interface~\cite{cuptiweb} provides CUDA host callback and device activity measurements.
Additionally, the hardware and operating system are monitored through an asynchronous measurement that involves the periodic or on-demand interrogation of the operating system, hardware states, or runtime states (e.g., CPU use,
resident set size, memory ``high water mark''). The NVIDIA Management Library interface~\cite{nvmlweb} provides periodic CUDA device monitoring to APEX.  In previous work~\cite{wei2021memory}, APEX was
extended to capture additional timers and counters related to CUDA device-to-device memory transfers, as well as tracking memory
consumption on both device and host when requested with the \texttt{cudaMalloc*} API calls.

\begin{figure*}
    \centering
    \subfloat[Task tree example.]{
    \includegraphics[width=0.45\textwidth]{figures/tasktree.2.dot.pdf}
    \label{fig:taskgraphs:tree}
    }
    \subfloat[Task graph example.]{
    \includegraphics[width=0.45\textwidth]{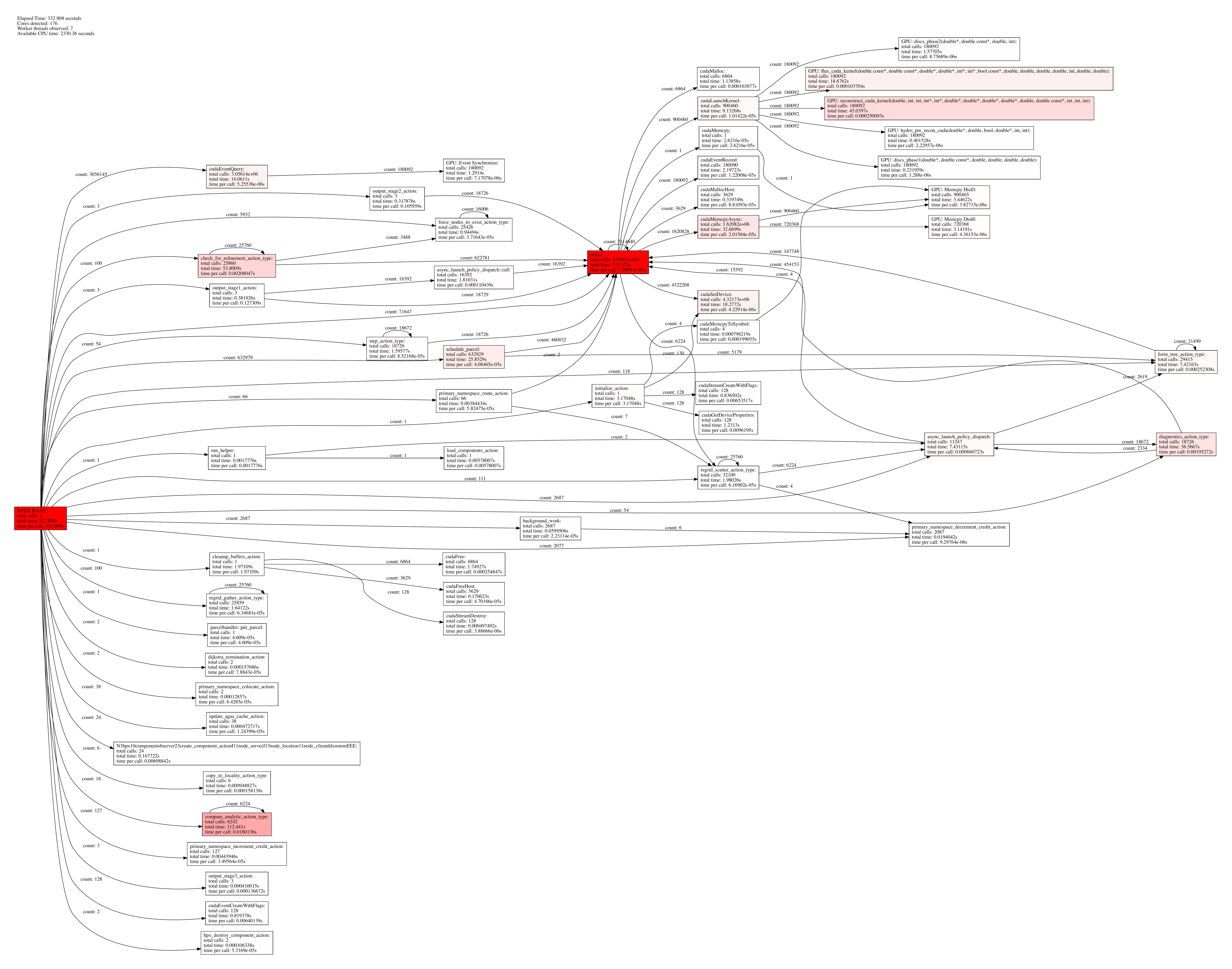}
    \label{fig:taskgraphs:graph}
    }
    \caption{Task tree and task graph of Octo-Tiger as captured by APEX.  Intensity of red color
    is correlated with the node's contribution to the overall runtime.  The recursive structure of the octree
    is evident in the expanded tree. High resolution images are available here (https://doi.org/10.6084/m9.figshare.14666184.v1).}
    \label{fig:taskgraphs}
\end{figure*}

%Updated support for counter scatterplot (to avoid OTF2 output)
Tracing measurement is typically used by application developers to understand timing and dependency
relationships between different tasks within an application.
When tracing to the Open Trace Format (OTF2) or Google Trace Events Format, each concurrent CUDA Stream is assigned three virtual ``threads" to
track kernel, memory and synchronization activity.  This is necessary because these three classes of events are not perfectly nested timers -- there is a potential for asynchronous overlap -- which are a requirement for the OTF2 tracing library (Google Trace Events are more forgiving).
However, each operation class within a Stream does
have a guaranteed ordering, so this segregation of event types is sufficient to meet the requirements of the tracing
libraries and formats.  However, because the Octo-Tiger CUDA implementation uses up to $128$ concurrent streams per process
(along with the actual HPX worker and helper threads on the CPU),
even a relatively small run with $6$ ranks per node can result in over 2400 unique ``threads" of execution, and
a collection of trace files over $27$GB in size from just $25$ iterations.
%Tasktree and taskgraph output
To work around this issue of scale, APEX was extended to support \textit{task dependency trees} to complement
the existing task dependency graph support.  The tree is a summary representation of the task dependency
relationships (\textit{task types, not individual tasks}), revealing the full dependency chain and not just immediate parent/child relationships.
While this can result in tree representations that are larger than the graph representation -- due to expanded
recursions and continuations, see Figure~\ref{fig:taskgraphs:graph} for an example -- the trees are still quite manageable and
helpful in diagnosing problems in
programming models like HPX that \textit{do not} have a meaningful callstack context but \textit{do} have
a task dependency context, including tasks and other activity offloaded to GPU devices.
%\todo[inline]{Maybe we should mention that each task(box) is a method that is being called more than once? Otherwise this graph looks like we aren't executing a lot of methods. Maybe we could add another subfigure which zooms in onto on of the boxes, showing the "number of calls" label? }
%(show examples?)
To complement the taskgraph and tasktree data in the absence of a full trace, APEX also captures task
and counter scatterplot data, indicating on the $x$ axis when the task started or the counter was captured, and
the $y$ axis contains the duration of the task or the value of the counter. The tasks are sampled using
a user-specified fraction (default 1\%) whereas the counters are sampled at every value capture.
This data collection allows the application developer to capture a time sequence of data without the
filesystem overhead of a full event trace.  Examples are shown in Section~\ref{sec:performance}.

%%%%%%%%%%%%%%%%%%%%%%%%%%%%%%%%%%%
\section{Octo-Tiger}
\label{sec:octo}
In this section, we briefly introduce the modules of Octo-Tiger studied in this paper, followed by details on how we integrate their GPU implementation with HPX. For a more general overview of the modules themselves, we refer to Octo-Tiger's method paper~\cite{marcello2021octo}.
\subsection{Octo-Tiger's Gravity Solver}
Octo-Tiger uses a fast multipole method (FMM) for solving the gravity ~\cite{Marcello2017}. This particular implementation of the FMM globally conserves both linear and angular momenta to machine precision, and, when coupled to the hydro-dynamics solver, also globally conserves energy to machine precision. The solver uses a third order multipole expansion. Its accuracy can be varied by adjusting the opening criterion, $\theta$. Lower values of the opening criterion lead to stricter multipole acceptance criteria, requiring that multipoles be further away to interact. This increases the solver's accuracy at the cost of more computation time.

%%%%%%%%%%%%%%%%%%%%%%%%%%%%%%%%%%%
\subsection{Octo-Tiger's Hydro Implementation}
\label{sec:octo:hydro}
Octo-Tiger solves the equations of hydrodynamics using a finite volume method. It evolves the mass density, three linear momenta, and gas energy on a rotating adaptive mesh refinement (AMR) mesh. The AMR mesh is based on an octree structure, with each node of the octree being either not refined at all or fully refined with eight sub-grids of twice the resolution as their parent. By default, each of those sub-grids consists of $8^3$ cells, however, this is adjustable at compile-time to allow for more finely refined sub-grids with more cells (for instance $16^3$). The evolved variables reside on the leaf sub-grids of the octree. It additionally evolves an entropy tracer, using it to implement the dual energy formalism of Bryan et al.~\cite{Bryan1995}. First, the evolution variables are reconstructed from cell averages at $26$ quadrature points on the cell face: the centers of cell faces and cell edges and at cell vertices. This is accomplished by applying the piece-wise parabolic method (PPM) of Colella et al.~\cite{Colella1984}. This third order, five cell stencil is applied along the lines between cell centers that coincide with particular quadrature points, producing left and right values for each. Octo-Tiger optionally allows for the contact discontinuity detection available with PPM. Once the evolution values are reconstructed, the semi-discrete central-upwind scheme of Kurganov et al.~\cite{kurganov2000} is applied to the reconstructed left and right variables at the quadrature points, producing fluxes. These fluxes are summed at quadrature points on a given cell face using Newtonian quadrature to obtain the final flux. Octo-Tiger's complete hydro scheme is described by Marcello, Shiber, et al.~\cite{marcello2021octo}. In this paper, we compare our new hydro module to the old hydro module. The old hydro module used the same reconstruction method, however, flux values were only computed at the centers of cell faces.

%%%%%%%%%%%%%%%%%%%%%%%%%%%%%%%%%%%
\subsection{Octo-Tiger's CUDA Implementation}
To understand Octo-Tiger's GPU implementation of the hydro module, it is worth reintroducing the GPU implementation of the gravity module from prior work.
While the gravity module uses entirely different compute methods (which we will only briefly mention here), it uses the same mechanism for combining HPX and CUDA to facilitate concurrent GPU kernel execution. The following subsection offers details how (and why) we use this mechanism, followed by the details of the hydro GPU implementation in the subsequent subsections.
\subsubsection{Gravity Module GPU Implementation}
The gravity solver---more specifically the calculation of the same-level interactions in the second FMM step---was the original hot spot within Octo-Tiger~\cite{Pfander18acceleratingFull, master_thesis_2018_daiss}.
Here, we have to calculate the cell-to-cell interactions for each of the cells of a sub-grid. The exact number of interactions per cell depends on the parameter $\theta$.
The actual hot spot consisted of different methods (henceforth called gravity kernels) that take care of the various types of cell-to-cell interactions. All kernels operate on one sub-grid at a time, calculating all interactions between the cells within that sub-grid in addition to their interactions with cells in the ghost layer. The interaction types and the gravity kernels themselves are detailed in prior work in more detail~\cite{Pfander18acceleratingFull}.

As a sub-grid only contains $512$ cells by default, a gravity kernel responsible for calculating the interactions of a single sub-grid does not cause enough work to saturate a GPU. % (as it has 512 workitems, each summing up all interactions for one cell). %one of those kernels does not present enough work to utilize a GPU. % ...
There are two ways to address this. As mentioned previously, the number of cells per sub-grid can be increased, which in turn would provide more work for each GPU kernel. However, this would be an Octo-Tiger specific solution.
Instead, we were previously able to overcome this limitation for the gravity-solver GPU kernels by using a more general approach: A HPX-CUDA integration. 

This integration allows for the execution of CUDA kernels to be integrated with the HPX runtime system via HPX futures.
Essentially, after launching a CUDA kernel, HPX offers the functionality to return a HPX future for it. The HPX scheduler will then continue to poll a CUDA event that will be set as soon as said CUDA kernel is done. Once the event is set, the HPX future will be set to ready, which in turn triggers all tasks that depend on it. This allows us to integrate CUDA kernels into the HPX task graph.

We can thus handle CUDA kernels (and CPU/GPU data transfers) the same way as any other HPX task, making it easily possible to chain them with other tasks, such as arbitrary CPU compute tasks, inter-node communication, or I/O.
Crucially, this means that the execution of a CUDA kernel gets automatically overlapped with other tasks, which includes the execution of other CUDA kernels on separate CUDA streams.
This leads to the concurrent execution of multiple CUDA kernels on separate sub-grids, preventing GPU starvation despite the small workload with just $512$ workitems per kernel invocation.

As we launch each CUDA kernel within a normal HPX task, we can easily suspend the task until the GPU kernel is done (as indicated by its HPX future) and have an HPX worker thread pick up the original task afterwards to process its results.
This allows a single worker thread to easily handle multiple CUDA streams, switching between HPX tasks.
In previous work, we achieved a high GPU utilization and performance using this approach within the gravity solver~\cite{daiss2019piz}.
There, we used $12$ worker threads (one for each CPU core) and $128$ CUDA streams for one P100 GPU.

% Considering the size of our kernels is dictated by the utilized size of the sub-grids, the overlapping of the CUDA kernels is especially important to utilize the GPU for the smaller sub-grids
For this approach, however, we need to keep any GPU-wide synchronization to a minimum.
This includes calls to \texttt{cudaMalloc()}
and the creation of CUDA streams.
To avoid creating more CUDA streams than necessary, we pre-allocate them at the start of the simulation. We usually use a pool of $128$ HPX CUDA executors per device, each handling one CUDA stream.
We further employ a GPU-buffer manager to avoid on-the-fly allocation of buffers as much as possible.
If available, the manager reuses previously allocated but currently unused device buffers from previous kernel invocations.
Only if none is available a new buffer will be created.

Both the HPX-CUDA integration (exposed with HPX futures) and the buffer manager (exposed by a set of allocators within the library CPPuddle) can now be used independent of Octo-Tiger, to allow a similar scheme of easy, task-based, concurrent GPU kernel execution in other applications.
This also means we can also easily re-use this technique to port more of Octo-Tiger's solvers to the GPU.

Furthermore, if needed, this CUDA-HPX integration approach can be combined with the other approach mentioned to increase GPU utilization: Increasing the size of the sub-grids. This allows us to approach the issue both on the tasking level using the integration and on the data-structure level by using sub-grids with more cells.

% Octo-Tiger utilizes a HPX CUDA integration which allows for the execution of CUDA kernels to be integrated with the HPX runtime system via HPX futures. This means we can handle CUDA kernels (and CPU/GPU data transfers) as we can any HPX task, making it easily possible to chain them together with arbitrary other tasks within Octo-Tiger, like for instance CPU kernel exeuction or inter-node communication. Crucially, this means that the execution of a CUDA kernel automatically gets overlapped with arbitrary other ongoing tasks, which includes the execution of other CUDA kernels on separate CUDA streams.
% Considering the size of our kernels is dictated by the utilized size of the sub-grids, the overlapping of the CUDA kernels is especially important to utilize the GPU for the smaller sub-grids.
% Previously, we only supported the GPU execution of most compute-intensive methods in the gravity solver as those were the hotspots of the application.
%Using the aforementioned HPX-CUDA integrations as well as pre-allocated GPU buffers and CUDA streams , we achieve good GPU performance here \cite{daiss2019piz}.

\subsubsection{Initial Hydro Module GPU Implementation}
\label{sec:octo:hydro:gpu}
Between the GPU implementation of the gravity module and the changes moving from the old hydro (where flux values are only computed at the centers of cell faces) to the new one as outlined in Section~\ref{sec:octo:hydro}, the hydro module becomes the new application hot spot.
Hence, we have ported the relevant methods of the hydro solver to CUDA for this work.
    %\item In this work we investigate the results of this first GPU implementation of the hydro solver.
The two major hot spots within the solver are the \texttt{reconstruct} method and the \texttt{compute\_fluxes} method (henceforth called hydro kernels).
The reconstruct method reconstructs the evolution variables using the PPM method as mentioned in Section~\ref{sec:octo:hydro}. In turn, the flux method takes care of computing the fluxes and the Newtonian quadrature to obtain the final flux.

Just as the kernel of the gravity solver, each hydro kernel operates on one sub-grid in each invocation.
Therefore, we are facing the same challenge as for the gravity solver: One kernel invocation on its own is insufficient to prevent GPU starvation.
We have therefore ported the hydro solver's methods into CUDA kernels in two steps:
First, we have optimized the kernels to run efficiently on a GPU. We have removed any excessive branching within the method (to avoid warp divergence), we have flattened all required data structures into one-dimensional arrays of continuous memory and removed any remaining, unnecessary memory in-directions of the initial CPU implementation.
Second, we have integrated the kernels into the HPX task graph as we did with the gravity kernel to facilitate concurrent GPU kernel execution and the overlap of data transfers.
\subsubsection{Next steps for the Hydro GPU Implementation}
While porting the hydro solver to CUDA resolves a major bottleneck within Octo-Tiger, the kernels themselves are still an initial implementation and thus not yet tuned to the maximum extent:
%However, before further optimizations,
We first need to evaluate whether the concurrent execution of the multiple GPU hydro kernels with several CUDA streams and HPX futures is sufficient for GPU utilization.
While we had achieved good results with this approach within the gravity solver~\cite{daiss2019piz}, the hydro kernels are less compute-intensive than the gravity kernels. %(where we have to compute up to 1074 interactions per workitem).%making it possible we are reaching the limits of this approach.
Thus, we might reach the limits of this approach. %Extremely short-running kernels (even compared to the gravity kernels)

If we do, there are multiple ways to address the issue:
The easiest way is to simply increase the size of the sub-grids, providing more work per kernel invocation, increasing the number of blocks in the CUDA launch configuration.
This makes it both easier to utilize the entire device and to increase the likelihood of having multiple resident blocks per SM which increases occupancy and thus hides latency.
Of course, a higher sub-grid size comes with the trade-off of decreased scalability as (given the same overall grid size) we have less sub-grids to distribute to the different compute nodes.
A more sustainable method would be to combine the kernels of multiple sub-grids into one kernel. However, this kind of work aggregation is more tedious to implement and comes with several implementation challenges of its own.

Thus, the current state of the CUDA implementation in this work provides a good starting point to evaluate the performance, before moving forward to fine-tuning the kernels themselves.
We have therefore enabled Octo-Tiger to be configured with larger sub-grid sizes at compilation time, and we will study its performance and scalability impact in the following sections.
A significant performance impact of larger sub-grid sizes in the hydro kernels would be a strong indication that we should focus on further work-aggregation before any fine-tuning of the compute kernels themselves.

%%%%%%%%%%%%%%%%%%%%%%%%%%%%%%%%%%%
\section{Performance measurements}
\label{sec:performance}
In this section, we examine the scaling of Octo-Tiger on ORNL's Summit. Table~\ref{tab:toolchain} shows the
toolchain that compiled Octo-Tiger. Table~\ref{tab:hardware} lists the hardware information of ORNL's Summit.
Note that we used 128 streams per V100. Disclaimer: Due to a testbed allocation on Summit, we had limited node
hours, which limited the possible performance measurements. In addition, for jobs with more than $128$ nodes we
experienced some error from the IBM\textregistered~Spectrum MPI on Summit that we send too many messages and a network device crashed, see IBM\textregistered~ticket TS005902510.
%Due to the challenge
%to debug large jobs, we could not show these scaling results here.
We therefore cannot show scaling results beyond $128$ nodes.
Strong scaling was used for all runs.

\begin{table}[tb]
    \centering
    \caption{Toolchain and Octo-Tiger's dependencies. }
    \begin{tabular}{ll|ll}\toprule
         gcc & 8.1.1/9.1.0 & hwloc & 1.11.12  \\
         spectrum-mpi & 10.3.1 & boost & 1.70.0  \\
         cuda & 11.2.0 & jemalloc & 5.1.0  \\
         hpx & 1.6.0 & silo & 4.10.2 \\
         hdf5 & 1.8.12 & cppuddle & \texttt{d32e50b} \\\bottomrule
    \end{tabular}
    \label{tab:toolchain}
\end{table}

\begin{table*}[tb]
    \centering
    \caption{ORNL's Summit hardware information}
    \begin{tabular}{ll|ll}\toprule
     GPUs & $6$ NVIDIA\textregistered~Volta\texttrademark~V$100$ & CPU  & $2$ IBM\textregistered~ POWER9\texttrademark \\
     OS &
RHEL 7.4 & Kernel & 4.14.0 \\
\multicolumn{2}{c|}{Interconnect} & \multicolumn{2}{c}{Mellanox\textregistered~EDR $100$G InfiniBand}\\\bottomrule
    \end{tabular}
    \label{tab:hardware}
\end{table*}

%%%%%%%%%%%%%%%%%%%%%%%%%%%%%%%%%%%
\subsection{Sedov-Taylor Blast Wave (Pure Hydro)}
To benchmark the new hydro kernels, the Sedov-Taylor blast wave is used. Table~\ref{tab:blast:data} shows the details of each level of refinement.
\begin{table}[tb]
    \centering
    \caption{Simulation details of the Sedov-Taylor blast wave. Note that each configuration has $16,777,216$ cells to be processed.}
    \begin{tabular}{ccc}\toprule
      Sub-Grid Size & Sub-Grid Count & Refinement level   \\
      $8^3$ & $32768$ & $5$ \\
      $16^3$ & $4096$ & $4$ \\
      $32^3$ & $512$ & $3$ \\\bottomrule
    \end{tabular}
    \label{tab:blast:data}
\end{table}

%%%%%%%%%%%%%%%%%%%%%%%%%%%%%%%%%%%
\subsubsection{Node level scaling}

The scaling on one Summit node is presented in this section. Each configuration with an increasing sub-grid size, see Table~\ref{tab:blast:data}, is executed on a single node using CPUs and CPUs + GPUs.
We start with one HPX locality, which is equivalent to one MPI process. Thus, using six HPX localities, we run six MPI processes on Summit. We chose this setup to enable easy multi-G{PU} usage, at the expense of more inter-process communication. For each HPX locality, we assigned seven CPU cores and none of the six GPUs.
Figure~\ref{fig:blast:node:level:cpu} shows the scaling with the increasing number of localities.

The CPU-only scaling for the sub-grid sizes of $8^3$ and $16^3$ behaves similarly, and the sub-grid size of $32^3$ performs better for three and more localities.

For the next run, one locality was assigned to seven CPU cores and one NVIDIA\textregistered~V$100$ GPU. With six localities, all available CPU cores and GPUs on a single node are utilized. We assigned $128$ CUDA streams to each locality. Note that for the sub-grid size of $32^3$ we had to decrease the number of streams for the run with one locality, since queuing too many large kernels caused the device to hit its memory limit.

Figure~\ref{fig:blast:node:level:gpu} shows the number of processed sub-grids per second. With increasing sub-grid size, the number of cells processed per second improves notably, even though the overall grid size stays the same (albeit consisting of fewer sub-grids).
As mentioned in Section~\ref{sec:octo:hydro:gpu}, the hydro GPU kernels might not offer enough work to prevent GPU starvation, even with running multiple kernels (on separate sub-grids using separate CUDA streams) concurrently on the GPU. Increasing the sub-grid size increases the amount of work per kernel accordingly, making it easier to scale up to an entire GPU simply by having more blocks of work items available. Of course, it also increases the chance of having multiple blocks resident on one SM (we ensure that register usage is low enough for multiple blocks to be resident on one SM during the compilation time), increasing occupancy and thus hiding latencies more efficiently.
The average runtime per \texttt{reconstruct} kernel is just $258$ microseconds, or $108$ microseconds for the \texttt{flux} kernel when using a sub-grid size of $8^3$, further highlighting this point.
%\todo[inline]{@Kevin: Are those average runtimes correct?}.
% Kevin says: yes, for the 8^3 case blast wave, those times are correct.
In the short term, we can offset this problem by using a larger sub-grid size. However, an explicit work aggregation scheme combining multiple sub-grids might be preferable as a long-term solution.

Overall, we get a reasonable speedup for using the GPUs given the initial state of our hydro implementation. For all sub-grid sizes, the processed number of sub-grids was one order of magnitude higher.
\begin{figure}[tb]
    \centering
    \subfloat[\label{fig:blast:node:level:cpu}]{
    \begin{tikzpicture}
    \begin{axis}[xlabel={\# localities},ylabel={Cells processed per second},title={\textbf{CPU only} (Pure Hydro)},grid,xtick=data,legend pos=north west]
    \addlegendimage{empty legend};
    \addplot[plot1,thick,mark=*] table [x expr=\thisrowno{0},y expr={16777216*10/\thisrowno{1}}, col sep=comma] {blast-8-node-cpu.csv};
    \addplot[plot2,thick,mark=*] table [x expr=\thisrowno{0},y expr={16777216*5/\thisrowno{1}}, col sep=comma] {blast-16-node-cpu.csv};
    \addplot[plot3,thick,mark=*] table [x expr=\thisrowno{0},y expr={16777216*5/\thisrowno{1}}, col sep=comma] {blast-32-node-cpu.csv};
    \legend{Sub-grid size,$8^3$,$16^3$,$32^3$};
    \end{axis}
    \end{tikzpicture}
    }

    \subfloat[\label{fig:blast:node:level:gpu}]{
    \begin{tikzpicture}
    \begin{axis}[xlabel={\# localities},ylabel={Cells processed per second},title={\textbf{CPU + GPU} (Pure Hydro)},grid,xtick=data,legend pos=north west]
    \addlegendimage{empty legend};
    \addplot[plot1,thick,mark=*] table [x expr=\thisrowno{0},y expr={16777216*25/\thisrowno{1}}, col sep=comma] {blast-distributed-8-node.csv};
    \addplot[plot2,thick,mark=*] table [x expr=\thisrowno{0},y expr={16777216*25/\thisrowno{1}}, col sep=comma] {blast-distributed-16-node.csv};
    \addplot[plot3,thick,mark=*] table [x expr=\thisrowno{0},y expr={16777216*25/\thisrowno{1}}, col sep=comma] {blast-distributed-32-node.csv};
    \legend{Sub-grid size,$8^3$,$16^3$,$32^3$};
    \end{axis}
    \end{tikzpicture}
    }
    \caption{Cells processed per second for the node level scaling. For one up to 6 localities on one Summit node. One locality was assigned to seven CPUs and one NVIDIA\textregistered~V$100$ GPU. }
    \label{fig:blast:node:level}
\end{figure}

%%%%%%%%%%%%%%%%%%%%%%%%%%%%%%%%%%%
\subsubsection{Distributed scaling}

The scaling up to $128$ Summit nodes using $768$ NVIDIA\textregistered~V$100$ GPUS and $5376$ CPU cores is studied. Here, we use $6$ localities with one GPU and $7$ CPU cores per node. Figure~\ref{fig:blast:distributed:grid} shows the number of sub-grids processed per second. Here, the sub-grid size of $16^3$ performs slightly better than the sub-grid size of $8^3$. For up to $8$ nodes the sub-grid size of $32^3$ performs best, but later not enough work is available, and the scaling flattens out. Figure~\ref{fig:blast:distributed:speedup} shows the speedup with respect to a single node. For up to $8$ nodes all sub-grid sizes perform similarly and the largest sub-grid size flattens out again. Up to $16$ nodes the lower two sub-grid sized perform similar and later the smallest sub-grid size performs best.

We need at least $7$ sub-grids per locality ($42$ per node), as otherwise the $7$ CPU cores are underutilized.
While the majority of the work is done by the GPUs, there are pre-processing steps and the procedure of sending the data to the GPU and launching the kernels that are done purely by the CPU.
Ideally, we have more sub-grids per locality, to truly benefit from the overlapping of computation, inter-locality communication and CPU/GPU data transfers that we gain by using the task-based functionality offered by HPX.
Indeed, we can observe good scaling as long as we have about $21$ sub-grids per locality, as we both have enough work for all cores and the GPU and benefit from the overlapping.
The parallel efficiency degrades visibly when going below that threshold. First, we start losing the benefits of overlapping. Later on, we simply cannot use all CPU cores of a locality to do the pre-processing, kernel launches and communication tasks (as one core always works on one sub-grid). Lastly, we hit the point where we only have one sub-grid per locality. Here, we naturally do not benefit at all by adding more nodes.

We can see this in the runs with sub-grid size $32^3$. Here we go below $21$ sub-grids per locality in-between $4$ and $8$ nodes (as we use $6$ localities per node), afterwards we go below $7$ sub-grids at $16$ nodes. Lastly, we hit $1$ sub-grid per locality at $64$ nodes, so further increasing the node count to $128$ makes no difference.

It is worth noting that the largest run with sub-grid size $8^3$ and $128$ nodes results in a runtime per timestep of just $286$ms, while with a sub-grid size of $16^3$ we get a runtime per timestep of $211$ms. Considering each timestep consists of three consecutive iterations of the hydro solver (due to Octo-Tiger's use of a third-order Runge Kutta time integration scheme) this highlights that even small inefficiencies and barriers could cause significant slowdowns, simply due to the short runtimes involved.

\begin{figure}[tb]
    \centering
    \subfloat[\label{fig:blast:distributed:grid}]{
    \begin{tikzpicture}
    \begin{axis}[xlabel={\# nodes},ylabel={Cells processed per second},grid,xtick={1,2,4,8,16,32,64,128,256,512,1024}, xmode=log,log basis x={2},legend pos=north west,title=Blast wave (Pure Hydro)]
    \addlegendimage{empty legend};
    \addplot[plot1,thick,mark=*] table [x expr=\thisrowno{0},y expr={16777216*25/\thisrowno{1}}, col sep=comma] {blast-distributed-8.csv};
    \addplot[plot2,thick,mark=*] table [x expr=\thisrowno{0},y expr={16777216*25/\thisrowno{1}}, col sep=comma] {blast-distributed-16.csv};
    \addplot[plot3,thick,mark=*] table [x expr=\thisrowno{0},y expr={16777216*25/\thisrowno{1}}, col sep=comma] {blast-distributed-32.csv};
    \legend{Sub-grid size,$8^3$,$16^3$,$32^3$};
    \end{axis}
    \end{tikzpicture}
}

 \subfloat[\label{fig:blast:distributed:speedup}]{
    \begin{tikzpicture}
    \begin{axis}[xlabel={\# nodes},ylabel={Speedup},grid,xtick={1,2,4,8,16,32,64,128,256,512,1024}, xmode=log,log basis x={2},legend pos=north west,ymode=log,log basis y={2},title=Blast wave (Pure Hydro)]
    \addlegendimage{empty legend};
    \addplot[plot1,thick,mark=*] table [x expr=\thisrowno{0},y expr={358.128/\thisrowno{1}}, col sep=comma] {blast-distributed-8.csv};
    \addplot[plot2,thick,mark=*] table [x expr=\thisrowno{0},y expr={118.317/\thisrowno{1}}, col sep=comma] {blast-distributed-16.csv};
    \addplot[plot3,thick,mark=*] table [x expr=\thisrowno{0},y expr={72.7865/\thisrowno{1}}, col sep=comma] {blast-distributed-32.csv};
    \addplot[domain=1:128]{x};
    \legend{Sub-grid size,$8^3$,$16^3$,$32^3$,Optimal};
    \end{axis}
\end{tikzpicture}

 }
    \caption{Cells processed per second for the distributed scaling from one Summit node up to 128 Summit nodes. Note that all six NVIDIA\textregistered~V$100$ GPUs per node were used.}
    \label{fig:blast:distributed}
\end{figure}

%%%%%%%%%%%%%%%%%%%%%%%%%%%%%%%%%%%
\subsection{Rotating star (Hydro and gravity)}
For the second example, the rotating star problem is studied, where the gravity solver is added to the hydro solver. Table~\ref{tab:star:data} shows the details for each level. Here, we use the default $\theta$ value ($0.5$) for the rotating star problem, which leads to fewer cell-to-cell interactions than we encounter with production run simulations.
This makes the gravity solver less compute-intensive than it would typically be. 
Furthermore, we have redesigned the gravity GPU kernels to allow different (larger) stencil sizes, making them currently less finely tuned than they previously were, as the shared-memory implementation in the monopole-monopole gravity kernel assumed a fixed stencil size.
Still, the rotating star scenario presents a good benchmark as it allows us to test the hydro- and gravity solver together in a simple scenario.

\begin{table*}[tb]
    \centering
    \caption{Simulation details of the rotating star. Note that each configuration has $16,777,216$ cells.}
    \begin{tabular}{cccc}\toprule
     Sub-Grid Size & Sub-Grid Count & AMR boundaries & Refinement level   \\
     $8^3$ & $44472$ & $3800$ & $8$ \\
     $16^3$ & $5944$ & $3800$ & $7$ \\\bottomrule
    \end{tabular}
    \label{tab:star:data}
\end{table*}

%%%%%%%%%%%%%%%%%%%%%%%%%%%%%%%%%%%
\subsubsection{Node level scaling}

%\todo[inline]{Merge itemize with text below}
%\begin{itemize}
%    \item Mention root node $N^2$ complexity (we need to handle all interactions that were not processed on lower level - it is basically on open stencil that only excludes interaction partners of a cell that are close and have thus been processed on a lower tree level already which) leads to performance degradation on CPU-only runs, as only one core works on the root node kernel.
%    \item On GPU runs this kernel will be run on the GPU, instead of a single POWER9\texttrademark core, which is a lot faster.
%    \item Overall for the gravity module, larger grid sizes are not beneficial as the gravity interactions kernels contain a lot more work (executing a larger stencil) per workitem than the hydro kernels.
%    \item We can still see some benefit from increasing the sub-grid size in the GPU runs, as the scenario uses both gravity- and hydro solver and the hydro solver works more efficiently with larger sub-grids. Furthermore, in the rotating start scenario $\Theta$ is set to $0.5$ \todo{check actual stencil size}, which yields a smaller stencil than is used for double white dwarf scenarios. Hence, the overall runtime impact of the gravity solver is smaller than usual.
%\end{itemize}
The scaling on one Summit node is presented in this section. Each configuration with an increasing sub-grid size, see Table~\ref{tab:star:data}, is executed on a single node using CPUs and CPUs + GPUs.
We start with one HPX locality, which is equivalent to one MPI process. Therefore, using six HPX localities, we run six MPI processes on Summit. For each HPX locality, we assigned seven CPU cores and none of the six GPUs. Figure~\ref{fig:star:node:level:cpu} shows the node level scaling from one up to $6$ localities for CPUs only. The smaller sub-grid sizes perform better using the CPUs only. We suspect that this is due to the gravity solver's handling of the root sub-grid within the octree: We have to process all same-level interactions within the sub-grid (as there is no higher level available that would take care of those interactions within the FMM algorithm). The runtime of calculating these interactions is $\mathcal{O}(N^2)$ with $N$ as the number of cells in the root sub-grid. In a CPU-only run, the root node is processed like any other sub-grid, meaning the same-level interactions are calculated within one HPX task; thus, only one CPU core is working on it, while all other cores take care of other tasks. This increases the runtime substantially while increasing the size of the sub-grids in particular, since the entire next top-down tree-traversals within the FMM algorithm depend on the results of the root sub-grid. With an increasing number of CPU cores, more of them will simply be idle whilst waiting on these results. When increasing the number of localities, the ratio of the root sub-grid's work to the work of the remaining sub-grids on the root locality increases, resulting in a higher load imbalance.

Figure~\ref{fig:star:node:level:gpu} shows the node level scaling adding one GPU to each locality. In that case, the GPU kernels benefit of the larger sub-grid size and larger sub-grid sizes performs better.
The issue with the root sub-grid is less severe here as the interactions are not being calculated by one CPU core alone, but instead by a GPU kernel. Between this improvement, and the general better runtime behavior of the hydro kernels when dealing with larger sub-grids, the performance improves when switching to a sub-grid size of $16^3$. However, the speedup is less severe than with the Sedov-Taylor blast wave scenario as the gravity GPU kernels do not seem to benefit from larger sub-grid sizes (even with the improved GPU kernel for the root sub-grid). Again, the processed sub-grids per second are one order of magnitude higher adding the GPUs.

\begin{figure}[tb]
    \centering
    \subfloat[\label{fig:star:node:level:cpu}]{
    \begin{tikzpicture}
    \begin{axis}[xlabel={\# localities},ylabel={Cells processed per second},title={\textbf{CPU only} (Hydro and gravity)},grid,xtick=data,legend pos=north west]
    \addlegendimage{empty legend};
    \addplot[plot1,thick,mark=*] table [x expr=\thisrowno{0},y expr={16777216*10/\thisrowno{1}}, col sep=comma] {star-8-node-cpu.csv};
    \addplot[plot2,thick,mark=*] table [x expr=\thisrowno{0},y expr={16777216*10/\thisrowno{1}}, col sep=comma] {star-16-node-cpu.csv};
    \legend{Sub-grid size,$8^3$,$16^3$};
    \end{axis}
    \end{tikzpicture}
    }

    \subfloat[\label{fig:star:node:level:gpu}]{
    \begin{tikzpicture}
    \begin{axis}[xlabel={\# localities},ylabel={Cells processed per second},title={\textbf{CPU + GPU} (Hydro and gravity)},grid,xtick=data,legend pos=north west]
    \addlegendimage{empty legend};
    \addplot[plot1,thick,mark=*] table [x expr=\thisrowno{0},y expr={16777216*10/\thisrowno{1}}, col sep=comma] {star-distributed-8-node.csv};
    \addplot[plot2,thick,mark=*] table [x expr=\thisrowno{0},y expr={16777216*10/\thisrowno{1}}, col sep=comma] {star-distributed-16-node.csv};
    \legend{Sub-grid size,$8^3$,$16^3$};
    \end{axis}
    \end{tikzpicture}
    }
    \caption{Cells processed per second for the node level scaling. For one up to 6 localities on one Summit node. One locality was assigned to seven CPUs and one NVIDIA\textregistered~V$100$ GPU. }
    \label{fig:star:node:level}
\end{figure}

%%%%%%%%%%%%%%%%%%%%%%%%%%%%%%%%%%%
\subsubsection{Distributed scaling}
We now study scaling on up to $128$ Summit nodes using $768$ NVIDIA\textregistered~V$100$ GPUS and $5376$ CPU cores. Here, we use $6$ localities with one GPU and $7$ CPU cores per node. Figure~\ref{fig:star:distributed:grid} shows the processed sub-grids per second for increasing number of nodes. Again, for the combined hydro and gravity simulation, the larger sub-grid sizes results in slightly better performance. Larger sub-grid sizes have less effect on the gravity solver and predominantly accelerate the hydro solver. Therefore, we observe a similar picture as for the hydro-only scenario. It is worth noting that the runtime per time step on $128$ nodes for the sub-grid size $8^3$ is $\approx 0.48$ seconds, and for sub-grid size $16^3$ it is $0.45$ seconds. Note that for each time step, Octo-Tiger solves $3$ hydro steps and 6 FMM steps (the gravitational potential as well as its time derivative appear in the source equations for the hydrodynamics). Here, the same argument is valid that we have good scaling as long as we have $21$ sub-grids per locality. This indicates that approximately 16 million cells are not enough work for $768$ GPUs.

\begin{figure}[tb]
    \centering
    \subfloat[\label{fig:star:distributed:grid}]{
    \begin{tikzpicture}
    \begin{axis}[xlabel={\# nodes},ylabel={Cells processed per second},grid,xtick={1,2,4,8,16,32,64,128,256,512,1024}, xmode=log,log basis x={2},legend pos=north west,title=Rotating star (Hydro and gravity)]
    \addlegendimage{empty legend};
    \addplot[plot1,thick,mark=*] table [x expr=\thisrowno{0},y expr={16777216*10/\thisrowno{1}}, col sep=comma] {star-distributed-8.csv};
    \addplot[plot2,thick,mark=*] table [x expr=\thisrowno{0},y expr={16777216*10/\thisrowno{1}}, col sep=comma] {star-distributed-16.csv};
    \legend{Sub-grid size,$8^3$,$16^3$};
    \end{axis}
    \end{tikzpicture}
}

 \subfloat[]{
    \begin{tikzpicture}
    \begin{axis}[xlabel={\# nodes},ylabel={Speedup},grid,xtick={1,2,4,8,16,32,64,128,256,512,1024}, xmode=log,log basis x={2},legend pos=north west,ymode=log,log basis y={2},title=Rotating star (Hydro and gravity)]
    \addlegendimage{empty legend};
    \addplot[plot1,thick,mark=*] table [x expr=\thisrowno{0},y expr={196.507/\thisrowno{1}}, col sep=comma] {star-distributed-8.csv};
    \addplot[plot2,thick,mark=*] table [x expr=\thisrowno{0},y expr={121.612/\thisrowno{1}}, col sep=comma] {star-distributed-16.csv};
    \addplot[domain=1:128]{x};
    \legend{Sub-grid size,$8^3$,$16^3$,Optimal};
    \end{axis}
\end{tikzpicture}

 }
    \caption{Cells processed per second for the distributed scaling from one Summit node up to 128 Summit nodes. Note that all six NVIDIA~\textregistered~V$100$ GPUs per node were used.}
    \label{fig:star:distributed}
\end{figure}

%%%%%%%%%%%%%%%%%%%%%%%%%%%%%%%%%%%
\subsection{APEX + CUDA}

The introduced overhead for the APEX CUDA measurements was about $30$ seconds for the run on a full single node which is $\approx 8.5\%$ of the total execution time. This is slightly more than using APEX without the CUDA counters where the overhead was around one percent~\cite{diehl2021performance}. This overhead is likely caused by excessive callback processing for some frequently called but short-lived CUDA functions. In fact, because the algorithms support the ability for each locality to schedule work on more than one GPU, the profiling showed that the function \texttt{cudaSetDevice} is called over $4,322,208$ times during a $332$ second run.  In addition, HPX uses polling to detect GPU activity completion instead of callbacks --- polling provides faster throughput --- and performing the polling requires $3,056,145$ calls to \texttt{cudaEventQuery}. These frequent, short calls are fine on their own, but there is an observed overhead in measuring them.

\begin{figure*}
    \centering
    % This also has a pdf, if we want a higher/variable resolution figure...
    \includegraphics[width=\textwidth]{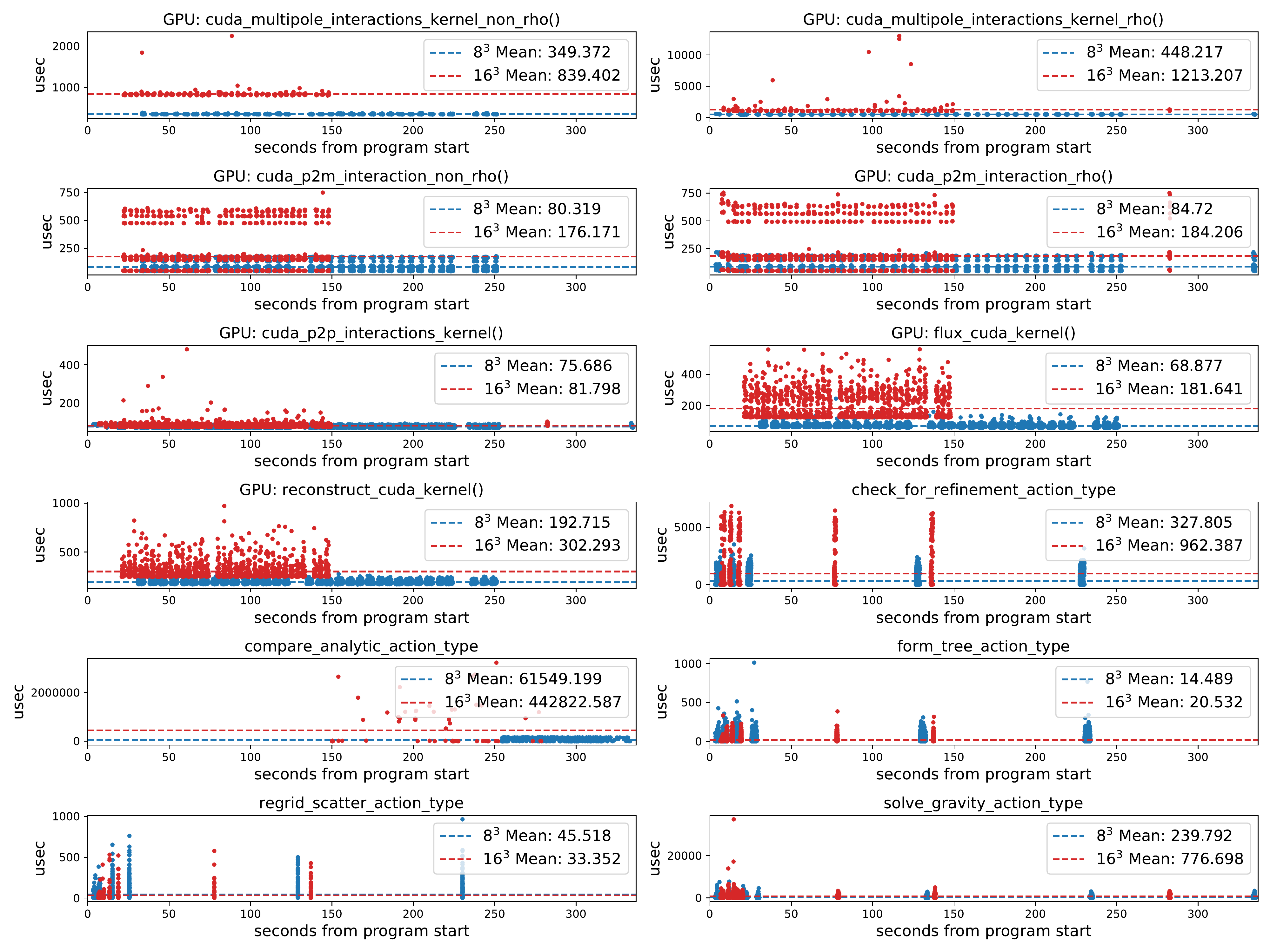}
    \caption{GPU kernel activity and CPU task actions for the gravity and hydro tasks when executing the $8^3$ and $16^3$ rotating star test on $6$ localities.  The $16^3$ decomposition leads to longer-running tasks and kernels, but a shorter overall execution time because there are significantly fewer of them.}
    \label{fig:kernel_activity}
\end{figure*}

Figure~\ref{fig:kernel_activity} shows the time spent in the sampled tasks during a short
execution of the rotating star problem.  The gravity (monopole/multipole interactions) and hydro
(flux\_cuda\_kernel, reconstruct\_cuda\_kernel) kernels execute on the GPU, whereas other actions
are executed on the CPU.  The validation routine (compare\_analytic\_action\_type) is executed on
the CPU only.  As this routine is only used for validating the results, it is unlikely to be ported
to the GPU.

\begin{figure}
    \centering
    \subfloat[CPU Utilization of the $168$ total available hardware threads.  Data captured by locality $0$ represents the aggregate utilization of all processes on the node.]{
    \label{fig:apex_counters:cpu}
    \includegraphics[width=0.47\textwidth]{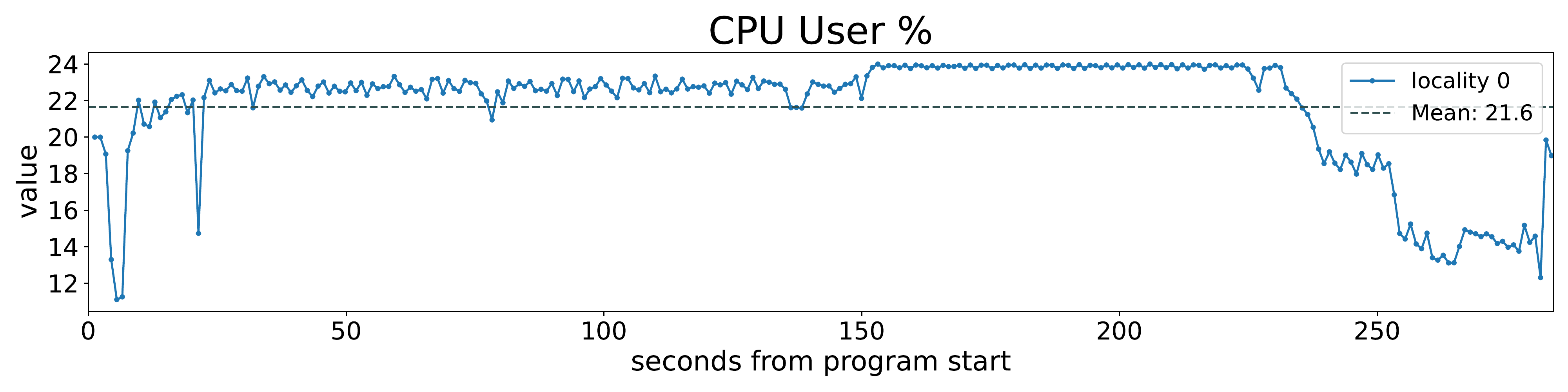}
    }\\
    \subfloat[GPU Utilization for Device $0$ used by locality $0$.]{
    \label{fig:apex_counters:gpu}
    \includegraphics[width=0.47\textwidth]{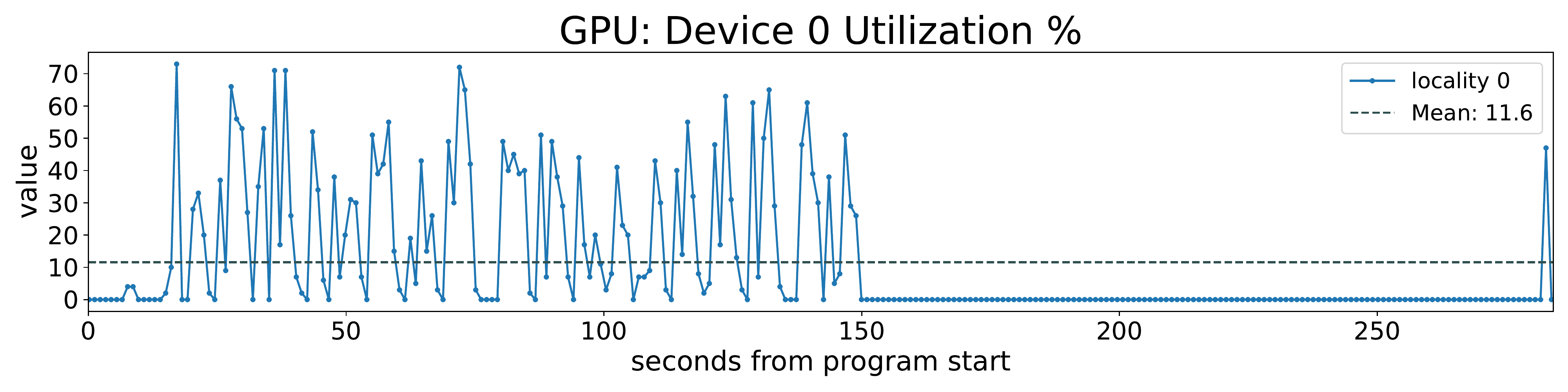}
    }\\
    \subfloat[Total memory occupied through explicit allocations on each GPU.]{
    \label{fig:apex_counters:gpu_mem}
    \includegraphics[width=0.47\textwidth]{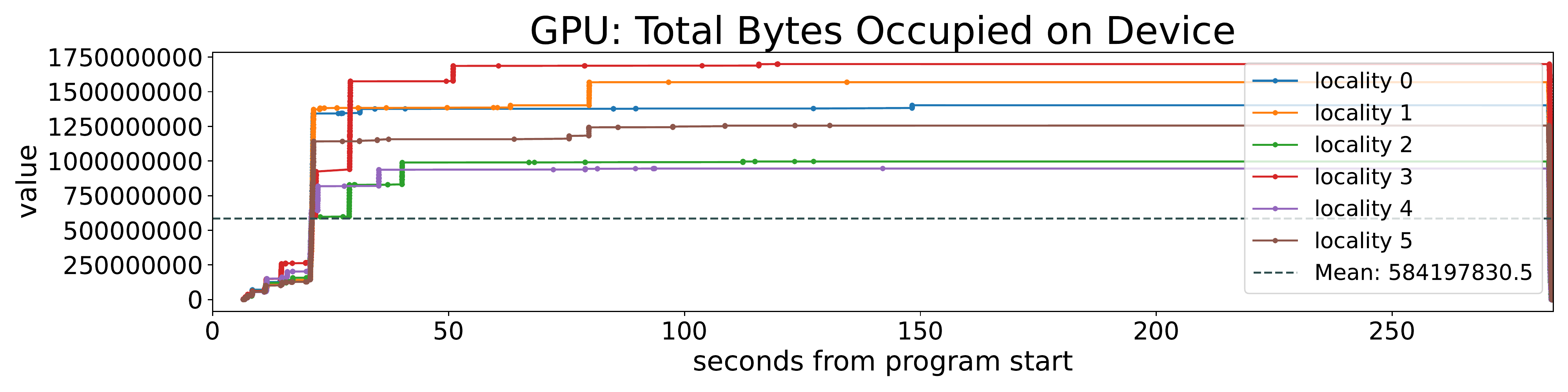}
    }
    \caption{APEX Performance counter metrics from the $16^3$ rotating star test case run on 6 localities.}
    \label{fig:apex_counters}
\end{figure}
%\todo[inline]{Considering the Slack comment about the GPU device utilization figure (7b) being inaccurate as it always looks at the same GPU (instead of all 6): Should we keep it or remove it?}

Figure~\ref{fig:apex_counters} shows three counters captured during the rotating star run that
indicate utilization of the allocated hardware.  The CPU user-space utilization in Figure~\ref{fig:apex_counters:cpu} is captured by monitoring the \texttt{/proc/stat} virtual file.  Although HPX has launched 1 worker thread per physical core,
the operating system detects 4 hardware threads per core.  Therefore, the maximum utilization possible in this configuration is $25\%$.  During the CPU-intensive validation at the end of execution, these threads are fully utilized, and during most of the execution the threads are well utilized.  Time spent processing system calls (not shown) peaks at $3\%$ during initialization and finalization and otherwise averages $0.66\%$. The GPU utilization data is captured by periodically capturing the available NVML data for device 0.  Finally, Figure \ref{fig:apex_counters:gpu_mem} shows the total memory allocated on the device through \texttt{cudaMalloc*()} calls, which peaks out at less than $11\%$ of available memory.  The GPU utilization and memory usage show that there is plenty of resources available to increase the amount of work per kernel and retain more data on the GPU.

%%%%%%%%%%%%%%%%%%%%%%%%%%%%%%%%%%%
\section{Astrophysical Test Results}
\label{sec:astro:results}
To verify that Octo-Tiger's new hydro module delivers better results for an equilibrium configuration, we ran a rotating star test problem. This star was constructed using a polytropic structural equation of state with the self-consistent field method (SCF)~\cite{Hachisu1986}. It is uniformly rotating about its $z$-axis at a rate sufficient to produce a star whose minor axis is $\sfrac{3}{4}$ the length of its major axis. We ran this problem for ten dynamical times. Since the star begins in equilibrium, we expect it to stay in equilibrium. We used two resolutions and for each resolution, two choices for the opening criterion, $\theta$. (Lower $\theta$'s result in a larger multi-pole interaction stencil for the gravity solver and hence better results). Here we define the density error as
\begin{equation}
    \rho_{L1} := \frac{\sum_\Omega (\rho_{IC} - \rho) \Delta^3}{V},
\end{equation}
where $\rho$ is the numerical mass density, $\rho_{IC}$ is the mass density from the initial conditions, $\Delta$ is a cell width, $V$ is the initial volume of the star, and the summation is over the entire domain $\Omega$. As shown in Table~\ref{tab:rotating}, in all cases the new hydro module delivers a lower error.

\begin{table*}[tb]
    \centering
\caption{The average error in the density field for the rotating star test using the old and new hydro modules. In these units, the central density of the star is $1$.}
    \begin{tabular}{cccc}\toprule
     Refinement Level & Opening Criterion & Old & New   \\
     $6$ & $0.5$ & $2.41\times 10^{-3}$ &$1.45\times 10^{-3}$\\
     $6$ & $0.35$ &$5.22\times 10^{-4}$ &$3.59\times 10^{-4}$\\
     $7$ & $0.5$ &$2.52\times 10^{-3}$ &$1.51\times 10^{-3}$\\
     $7$ & $0.35$ &$4.49\times 10^{-4}$ &$2.78\times 10^{-4}$\\\bottomrule
    \end{tabular}
    \label{tab:rotating}
\end{table*}

\section{Conclusion}
\label{sec:sonclusion}
This paper showed the following aspects in evaluating Octo-Tiger's performance on Summit. First, from the astrophysical aspect, the new implementation of the hydro kernel using a fully three-dimensional reconstruction of the fluxes is more computationally expensive than the old kernel. However, the new hydro kernel evolves an equilibrium rotating star with greater accuracy than the old kernel.

Second, the scaling on Summit showed the following two things. First, on a single node, the usage of the GPUs improved the cells processed per second by an order of magnitude. Thus, Octo-Tiger benefits from the usage of GPUs for the hydro, and combined hydro and gravity simulations. Second, the distributed scaling up to 128 nodes using 768 NVIDIA\textregistered~V$100$ GPUS and $5376$ CPU cores was presented. Both test problems scaled up to $128$ nodes for the two lower sub-grid sizes. However, we have seen that a problem containing $16,777,216$ cells starts to flatten out up to 128 nodes and indicates that even larger problems are necessary to provide enough work for the additional GPUs. With our testbed allocation on Summit, we could only show preliminary scaling results; however, we will continue to work to get the larger node counts running.

Third, the variation of sub-grid sizes was added to Octo-Tiger and this work studied the performance impact for the first time. For the hydro module on a single node, the sub-grid size of $32^3$ showed the best performance for the combined CPU and GPU runs, since with the larger sub-grid size more work was available for a single kernel run. However, for the distributed runs, only up to $8$ nodes the largest sub-grid size gave the best performance. For the combined hydro and gravity simulation, the sub-grid size of $16^3$ gives slightly better performance. This indicates that this sub-grid size will be the default for production runs.

Finally, the APEX CUDA profiling provides combined task trees and task graphs for the work on the GPU and CPU. Previously, Octo-Tiger was run first to profile the CPU usage with APEX and a second time with NVIDIA\textregistered profiler. The new plots provide some insights into the asynchronicity of HPX and the dependency of tasks. The scatter plots showed that the memory usage on the GPU was small, since only the data to be computed are kept in the device memory. In addition, we could show a good utilization of the CUDA devices on a single node. These plots provide a good base to analyze the combined asynchronous tasks on the CPU and GPU and support our efforts to optimize the concurrent CPU and GPU tasks.

\subsection{Future Work}
The results of this work motivate further improvements of the hydro solver's GPU implementation.
We plan to investigate on-the-fly work aggregation across sub-grids to combine the benefits of larger GPU kernels to saturate GPUs with the increased scalability that smaller sub-grids offer.

Furthermore, after recent promising results using HPX and Kokkos together within the gravity solver, we plan to port the current hydro CUDA implementation to Kokkos~\cite{CarterEdwards20143202} as well.
The HPX Kokkos integration works similarly as the CUDA one, and transforming the hydro CPU methods into GPU Kokkos kernels would have required the same changes to the methods themselves as outlined in Section~\ref{sec:octo:hydro:gpu}.
Hence, as of the current state, we have already completed the first important steps.%the porting the hydro kernel from CUDA to Kokkos is still straightforward.

Using Kokkos rather than pure CUDA provides us with two advantages: We can easily target GPUs of other vendors, such as AMD GPUs (and with the recently introduced Kokkos SYCL execution space, also Intel GPUs).
Furthermore, Kokkos provides the means of using explicit SIMD vectorization~\cite{sahasrabudhe2019portable} to run GPU-capable kernels efficiently on the CPU as well. Currently, we have to maintain a second set of CPU kernels using Vc for SIMD vectorization, which would be replaced by the Kokkos kernels. With a portable Kokkos implementation, there would be no need to maintain two specialized CPU and GPU kernels to cover all platforms anymore.  APEX already supports Kokkos profiling.
%\todo[inline]{cite IPDPSW paper?}

Furthermore, we plan to optimize the hydro kernels for shared memory usage as soon as they have been ported to Kokkos. With respect to HPX, more debugging is needed for jobs with larger node counts ($\geq$ $128$ nodes): We have experienced stalls for higher node counts due to an error from the IBM\textregistered~Spectrum MPI on Summit possibly caused by sending too many messages which result in a network device crash, see IBM\textregistered~ticket TS005902510.

%since we experienced consistent hanging at the initialization of the HPX run time. Since we never experienced this behavior using the openmpi/MPICH compiler wrappers on x86 and cray machines, it might relate to the IBM\textregistered~SPECTRUM MPI compiler wrapper.

From the application perspective, the authors would like to compare the performance of the rotating star with the Castro code to gain insight into whether the more accurate hydro module results in more stable shapes of the star. However, a comparison of the scaling is not trivial since different algorithms and solvers are used in both codes. In addition, Octo-Tiger utilizes asynchronous computation with HPX, which CASTRO does not, as it uses MPI+X. Next, these scaling results are the preparation for large production runs on GPU accelerated supercomputers.

% conference papers do not normally have an appendix
% use section* for acknowledgment
\footnotesize
\section*{Acknowledgment}
This research used resources of the Oak Ridge Leadership Computing
Facility, which is a DOE Office of Science User Facility supported under Contract DE-AC05-00OR22725. Diehl and Marcello thank the LSU Center of Computation \& Technology for supporting this work.
The APEX work was supported by the Scientific Discovery through Advanced Computing (SciDAC) program funded by U.S. Department of Energy, Office of Science, Advanced Scientific Computing Research (ASCR) under contract DE-SC0021299.

\section*{Supplementary materials}
The scripts to compile Octo-Tiger are available on GitHub~\cite{PowerTiger} and the script to run the jobs and the input files on Zenodo~\cite{patrick_diehl_2021_4777149}, respectively. CPPuddle is available here\footnote{\url{https://github.com/SC-SGS/CPPuddle}}.

\section*{Copyright notice}
\textcopyright 2021 IEEE. Personal use of this material is permitted.
  Permission from IEEE must be obtained for all other uses, in any current or future 
  media, including reprinting/republishing this material for advertising or promotional 
  purposes, creating new collective works, for resale or redistribution to servers or 
  lists, or reuse of any copyrighted component of this work in other works. 

\normalsize

% trigger a \newpage just before the given reference
% number - used to balance the columns on the last page
% adjust value as needed - may need to be readjusted if
% the document is modified later
%\IEEEtriggeratref{8}
% The "triggered" command can be changed if desired:
%\IEEEtriggercmd{\enlargethispage{-5in}}

% references section

% can use a bibliography generated by BibTeX as a .bbl file
% BibTeX documentation can be easily obtained at:
% http://mirror.ctan.org/biblio/bibtex/contrib/doc/
% The IEEEtran BibTeX style support page is at:
% http://www.michaelshell.org/tex/ieeetran/bibtex/
%\bibliographystyle{IEEEtran}
% argument is your BibTeX string definitions and bibliography database(s)
%\bibliography{IEEEabrv,../bib/paper}
%
% <OR> manually copy in the resultant .bbl file
% set second argument of \begin to the number of references
% (used to reserve space for the reference number labels box)
%\begin{thebibliography}{1}
%
%\bibitem{IEEEhowto:kopka}
%H.~Kopka and P.~W. Daly, \emph{A Guide to \LaTeX}, 3rd~ed.\hskip 1em %plus
%  0.5em minus 0.4em\relax Harlow, England: Addison-Wesley, 1999.
%
%\end{thebibliography}

\bibliographystyle{IEEEtran}
\bibliography{refs}

% that's all folks
\end{document}